\documentclass[prb,twocolumn,superscriptaddress,letterpaper]{revtex4}
\input epsf.sty
\usepackage{fullpage}
\usepackage{graphics}
\usepackage{fullpage}
\usepackage{graphicx}
\usepackage{epstopdf}
\begin{document}
\newtheorem{theorem}{Theorem}
\newtheorem{acknowledgement}[theorem]{Acknowledgment}
\newtheorem{algorithm}[theorem]{Algorithm}
\newtheorem{axiom}[theorem]{Axiom}
\newtheorem{claim}[theorem]{Claim}
\newtheorem{conclusion}[theorem]{Conclusion}
\newtheorem{condition}[theorem]{Condition}
\newtheorem{conjecture}[theorem]{Conjecture}
\newtheorem{corollary}[theorem]{Corollary}
\newtheorem{criterion}[theorem]{Criterion}
\newtheorem{definition}[theorem]{Definition}
\newtheorem{example}[theorem]{Example}
\newtheorem{exercise}[theorem]{Exercise} 
\newtheorem{lemma}[theorem]{Lemma}
\newtheorem{notation}[theorem]{Notation}
\newtheorem{problem}[theorem]{Problem}
\newtheorem{proposition}[theorem]{Proposition}
\newtheorem{remark}[theorem]{Remark}
\newtheorem{solution}[theorem]{Solution}
\newtheorem{summary}[theorem]{Summary}   
\def\r{{\bf{r}}}
\def\j{{\bf{j}}}
\def\m{{\bf{m}}}
\def\k{{\bf{k}}}
\def\kt{{\tilde{\k}}}
\def\mt{{\hat{t}}}
\def\mG{{\hat{G}}}
\def\mg{{\hat{g}}}
\def\mGa{{\hat{\Gamma}}}
\def\mS{{\hat{\Sigma}}}
\def\mT{{\hat{T}}}
\def\K{{\bf{K}}}
\def\P{{\bf{P}}}
\def\q{{\bf{q}}}
\def\Q{{\bf{Q}}}
\def\p{{\bf{p}}}
\def\x{{\bf{x}}}
\def\X{{\bf{X}}}
\def\Y{{\bf{Y}}}
\def\F{{\bf{F}}}
\def\G{{\bf{G}}}
\def\bG{{\bar{G}}}
\def\mbG{{\hat{\bar{G}}}}
\def\M{{\bf{M}}}
\def\V{\cal V}
\def\tchi{\tilde{\chi}}
\def\tx{\tilde{\bf{x}}}
\def\tk{\tilde{\bf{k}}}
\def\tK{\tilde{\bf{K}}}
\def\tq{\tilde{\bf{q}}}
\def\tQ{\tilde{\bf{Q}}}
\def\si{\sigma}
\def\ep{\epsilon}
\def\hep{{\hat{\epsilon}}}
\def\al{\alpha}
\def\be{\beta}
\def\ep{\epsilon}
\def\bep{\bar{\epsilon}_\K}
\def\up{\uparrow}
\def\de{\delta}
\def\De{\Delta}
\def\up{\uparrow}
\def\dwn{\downarrow}
\def\ksi{\xi}
\def\etha{\eta}
\def\product{\prod}
\def\goto{\rightarrow}
\def\switch{\leftrightarrow}
\title{Triplet Excitations in Carbon Nanostructures}
\author{K.~Aryanpour} 
\affiliation{Department of Physics, University of Arizona, Tucson, Arizona 85721, USA}
\author{S.~Mazumdar} 
\affiliation{Department of Physics, University of Arizona, Tucson, Arizona 85721, USA}
\affiliation{College of Optical Sciences, University of Arizona, Tucson, Arizona 85721, USA}
\author{H.~Zhao} 
\affiliation{School of Physics and Telecommunication Engineering, South China Normal University, Guangzhou, China}
\date{\today}
\begin{abstract} 
We show that the energy differences between the lowest optical singlet exciton and the lowest triplet exciton in semiconducting single-walled carbon nanotubes with diameter $\sim 1$ nm and graphene nanoribbons with widths $\sim 2$ nm are an order of magnitude smaller than in the $\pi$-conjugated polymer poly(para-phenylenevinylene). Our calculated energy gaps between the singlet and triplet excitons are in excellent agreement with the measured values in three different nanotubes with diameters close to 1 nm. The spatial extent of the triplet exciton is nearly the same as that of the singlet exciton in wide nanotubes and nanoribbons, in contrast to that in $\pi$-conjugated polymers, in which the triplet exciton exhibits strong spatial confinement. Weakly confined behavior of the triplet state begins in nanoribbons with widths as narrow as 2.5 times the graphene unit lattice vector. We discuss possible consequences of the small singlet-triplet energy difference in the carbon nanostructures on device applications.
\end{abstract}

\maketitle 
\section{Introduction}
\label{Intro}
\par Although the spin-singlet excitations in semiconducting single-walled carbon nanotubes (S-SWCNTs) have been widely investigated both theoretically and experimentally \cite{JorioBook}, the literature on spin-triplet excitations in these systems is relatively sparse. Theoretical predictions of the energy difference between the optically bright spin-singlet exciton and the lowest triplet exciton (hereafter $\Delta_{\mathrm{ST}}$) in the S-SWCNTs vary widely, with estimates of $\Delta_{\mathrm{ST}}$ in S-SWCNTs with diameter $d\sim1$ nm ranging from $\sim 20-40$ \cite{Perebeinos05a,Capaz07a} to $\sim 300$ meV \cite{Tretiak07a}. The earliest determination of spin triplet excitations were for S-SWCNTs with large $d \sim 1.4-1.8$ nm, where $\Delta_{\mathrm{ST}}$ are rather small \cite{Mohite09a} and precise calculations are difficult. Experimental identification of triplet excitations is made difficult by the occurrence of optically dark spin-singlet excitons below the singlet bright exciton \cite{Mortimer07a,Mortimer07b,Shaver07a,Kiowski07a,Srivastava08a,Matsunaga08a,Matsunaga09a}. It is significant, however, that early experimental work on S-SWCNTs with $d\sim1$ nm found two distinct groups of optically dark excitons \cite{Kiowski07a}, redshifted relative to the singlet bright exciton by $\sim 40$ meV and 100-130 meV, respectively, a point we come back to later. More recent careful measurements have determined the energy locations of the lowest triplet excitons in several S-SWCNTs with diameters close to 1 nm \cite{Harutyunyan09a,Matsunaga10a,Nagatsu10a}. The experimental $\Delta_{\mathrm{ST}}$ are different from both sets of theoretical predictions and are intermediate in magnitude (see below). In contrast to nanotubes, where considerable photophysical studies have been carried out, the existing literature on the photophysics of graphene nanoribbons (GNRs) is largely theoretical in nature \cite{Yang07a,Yang07b,Prezzi08a,Liao08a,Gundra11a}. Discussions here have been limited to spin-singlet excitations only. 
\par The sparseness of the literature on triplet excitations is surprising, since even though the spin-selection rule requires that optical excitations from the singlet ground state occur only to singlet excited states, triplet excitations can play strong indirect roles in optoelectronic applications. In the context of light emitting diodes with molecular or polymeric $\pi$-conjugated systems as the active materials, for example, the relative yields of light emissive singlet excitons versus nonemissive triplet excitons in electroluminescence is a topic of strong interest \cite{Cao99a,Kim00a,Wohlgenannt01a,Tandon03a,Dhoot02a,Segal03a,Carvelli11a}. The formation rates of the singlet and triplet excitons depend strongly on $\Delta_{\mathrm{ST}}$ \cite{Wohlgenannt01a,Tandon03a}. Similar discussions are ongoing also in the area of organic photovoltaics; the possibility of harvesting longlived triplet excitons that are reached from the singlet optical exciton via {\it intersystem crossing} (ISC) \cite{Arif09a} or {\it singlet fission} \cite{Smith10a} are being actively investigated. The efficiencies of both ISC and fission depend on $\Delta_{\mathrm{ST}}$. Thorough investigation of $\Delta_{\mathrm{ST}}$ and the nature of the spin-triplet excitations in the carbon nanostructures is thus called for. 
\par In the present paper, we report the results of theoretical calculations of triplet excitations in several S-SWCNTs and armchair GNRs (AGNRs) within a molecular correlated-electron model. We compare these results with those for poly(para-phenylenevinylene) (PPV) for which experimental results are known \cite{Monkman99a}. Our findings reveal that there is a {\it qualitative} change in the triplet state wave function in AGNRs with increasing width; nanoribbons with widths as narrow as 2.5 times the graphene lattice vector already exhibit small $\Delta_{\mathrm{ST}}$. Coulomb correlation effects on the honeycomb lattice appear to be different from those in one-dimensional chainlike systems. Similar behavior as AGNRs for the triplet state wave function is also found in S-SWCNTs 
with increasing diameter. 
\section{Model and Parameters}
\label{Model}
\par Our studies are within the correlated $\pi$ electron  Pariser-Parr-Pople (PPP) Hamiltonian \cite{Pariser53a,Pople53a}. For PPV, accurate theoretical results for both singlet and triplet excitations have previously been obtained within the PPP Hamiltonian \cite{Chandross97a}. The PPP model has also been used previously to investigate the spin singlet subspace in S-SWCNTs. For the longitudinal singlet excitons seen in optical absorption measurements with light polarized parallel to the nanotube axes, nearly quantitative fits to experimental absolute exciton energies and exciton binding energies have been obtained for nanotubes with diameter $d\geq0.75$ nm \cite{Wang06a}. Quantitatively accurate energies have also been obtained for the transverse excitons seen in absorption experiments with light polarized transverse to the nanotube axes \cite{Wang07a}. Similar accuracies are expected for the spin triplet subspace within the model.
\par We write the PPP Hamiltonian as,
\begin{eqnarray}
\label{Ham}
H = -\sum_{\langle ij \rangle, \sigma}t_{ij}
(c_{i,\sigma}^\dagger c_{j,\sigma}+ \mathrm{H.c.}) + 
 \sum_{i} Un_{i,\uparrow}n_{i,\downarrow} \nonumber \\ + \sum_{i<j} V_{ij} (n_{i}-1)(n_{j}-1). \hspace{0.5in}
\end{eqnarray}
where $c_{i,\sigma}^{\dagger}$ creates a $\pi$ electron of spin $\sigma$ on carbon atom $i$, $n_{\mu,\sigma} =  c_{i,\sigma}^{\dagger}c_{i,\sigma}$, and $n_{i} = \sum_{\sigma}n_{i,\sigma}$. We restrict the one-electron hopping $t_{ij}$ to nearest neighbors without loss of generality \cite{Wang07a}. $U$ is the repulsion between two electrons occupying the same $p_{z}$ orbital of a C atom and $V_{ij}$ are intersite Coulomb interactions.
\par Our choice of the $t_{ij}$ is system dependent. For PPV, we assume a planar configuration and choose $t_{ij}=2.4$ eV for phenyl C-C bonds and 2.2 (2.6) eV for the single (double) C-C bonds of the vinylene group \cite{Chandross97a}. We assume hydrogenation of the edge carbon atoms in AGNRs and take uniform $t_{ij}=2.4$ eV. In real systems, the site energies of the edge carbon atoms and hopping integrals linking them are expected to be slightly different from the corresponding quantities for the carbon atoms near the centers of the nanoribbons. However, we expect this effect to contribute weakly to the relatively large difference in $\Delta_{\mathrm{ST}}$ between PPV and AGNRs (see below). For the S-SWCNTs, based on earlier work, \cite{Wang06a} we set $t_{ij}=2.0$ eV. The curved nature of the S-SWCNT surface calls for the smaller value \cite{Wang06a}. 
\par The onsite Coulomb interaction $U$ is obviously identical for all three families. The intersite Coulomb interactions are parametrized as $V_{ij} = U/(\kappa\sqrt{1+0.6117 R_{ij}^2})$, where $R_{ij}$ is the distance between C atoms $i$ and $j$ in Angstroms. In the above expression, $\kappa=1$, with $U=11.26$ eV correspond to the standard Ohno parametrization \cite{Ohno64a} for the PPP model. Considerably better fits to experiments are obtained with slightly smaller $U$ and large $\kappa$. Based on the fitting of ground state as well as excited state absorptions in PPV \cite{Chandross97a} and S-SWCNTs, \cite{Wang06a,Wang07a} we chose $U=8.0$ eV and $\kappa=2$.
\par The model systems we have chosen for our theoretical study are (i) a 20-unit PPV chain capped at both ends, (ii) S-SWCNTs characterized by indices $(n,m)=(8,0)$, $(10,0)$, $(11,0)$, $(6,4)$, $(6,5)$ $(7,5)$, $(7,6)$, and $(9,4)$, and AGNRs, which we describe with the labeling scheme of Ezawa \cite{Ezawa06a}: $(p,q)=(15,1)$, $(19,1)$, and $(21,1)$. The zigzag S-SWCNTs were chosen because of the simplicity of performing calculations for these with a very large number of unit cells. Previous experience has shown that energy difference in nanotubes depend primarily on diameters $d$ and relatively weakly on chiralities. Thus the calculated $\Delta_{\mathrm{ST}}$ for the zigzag nanotubes may be considered representative for all nanotubes with diameters that are close. In contrast to the zigzag S-SWCNTs, experimental results for $\Delta_{\mathrm{ST}}$ are available for the $(6,4)$, $(7,5)$, and $(9,4)$ nanotubes. We show below that our calculated results for both the spin-singlet bright exciton and the lowest spin-triplet exciton are in excellent agreement with experiments in these cases. While the triplet subspace of the two remaining chiral S-SWCNTs in our list, $(6,5)$ and $(7,6)$, have not been studied experimentally yet, their photophysics has been widely investigated in the past. The long exciton lifetime in the $(6,5)$ S-SWCNT has allowed experimental estimation of the size and the mobility of the optical singlet exciton \cite{Luer08a}. Our calculated singlet exciton size compares very favorably with the experimental value. Our predicted theoretical results for triplet excitations can thus be compared against future experiments in the $(6,5)$ and $(7,6)$ S-SWCNTs. The AGNRs we have chosen are obtained by unrolling our zigzag nanotubes after cutting them along their translational vectors. 
\par As in previous works \cite{Wang06a,Wang07a}, our calculations for the S-SWCNTs and AGNRs are done with open boundary condition. For the zigzag S-SWCNTs, our calculations are for 50 unit cells with $\sim$ 2000 carbon atoms in each case. For the chiral S-SWCNTs the number of unit cells we retain is smaller, but the number of carbon atoms is comparable; in all cases, we retain larger than 2000 carbon atoms. Furthermore, for each of the three S-SWCNTs for which experimental results are available, we have done careful finite-size analysis to confirm that our calculated energies have converged to within a few percent. Calculations for the AGNRs are for 40 unit cells. 
\par Tractable approximate approaches to the many-body PPP Hamiltonian  are necessary to determine the excitation energies of large systems. As discussed in the context of singlet excitations \cite{Wang06a,Wang07a}, we use the single configuration interaction (SCI) approximation, which involves diagonalization of the PPP Hamiltonian in the space of single excitations from the Hartree-Fock (HF) ground state. The excitation of a single electron from the ground state can give either the singlet or triplet electron-hole pair or exciton. Indirect exchange within Eq.~(\ref{Ham}) differentiates between the two spin states.
\section{Results and Discussions}
\label{Results}
\par In Table~\ref{table1}, we have given our calculated singlet and triplet exiton energies (hereafter $E_{S}$ and $E_{T}$, respectively). In the case of the S-SWCNTS, there occur singlet dark excitons below the optical bright exciton. We have found that similar structures, with several close-lying triplet states, occur also in the spin triplet subspace. In Table~\ref{table1}, $E_{S}$ corresponds to the singlet bright exciton and $E_{T}$ to the {\it lowest} triplet exciton. Thus in S-SWCNTS, there occur several dark singlet and triplet excitons between these two states. We have also included in Table~\ref{table1} the calculated singlet exciton binding energies (hereafter $E_{bS}$), corresponding to the difference in energy between the lowest HF continuum within SCI and the singlet exciton \cite{Wang06a}. Quantitative agreement between the calculated $E_{S}$ and $E_{bS}$, and their experimental estimates \cite{Dukovic05a,Zhao05a,Bachilo02a} in S-SWCNTs have been pointed out earlier \cite{Wang06a}. We therefore compare theory and experiments here only for the S-SWCNTs for which triplet energies are available. For PPV, the calculated $E_{S}=2.69$, $E_{T}=1.46$, and $\Delta_{\mathrm{ST}}=1.23$ eV are to be compared against experimental quantities $2.4$, $1.3$, and $1.1$ eV, respectively \cite{Monkman99a}. The experimental triplet exciton energies for the $(6,4)$, $(7,5)$, $(9,4)$ S-SWCNTs in Table I were obtained from emissions from triplet states that became optically active upon pulsed-laser irradiation \cite{Harutyunyan09a,Matsunaga10a} or hydrogen absorption \cite{Nagatsu10a}. It is reasonable to assume that the emissions are from the lowest triplet exciton within the triplet manifold \cite{Harutyunyan09a,Matsunaga10a,Nagatsu10a}. Excellent agreement between theory and experiments is seen in all three cases. The most striking aspect of the results in Table~\ref{table1} is the very small $\Delta_{\mathrm{ST}}$ in the S-SWCNTs and the AGNRs, nearly one order of magnitude smaller than that in PPV. We do not find the simple $A/d^2$ diameter dependence of $\Delta_{\mathrm{ST}}$ suggested previously \cite{Capaz07a}.
\begin{table} 
\includegraphics[width=3.2in]{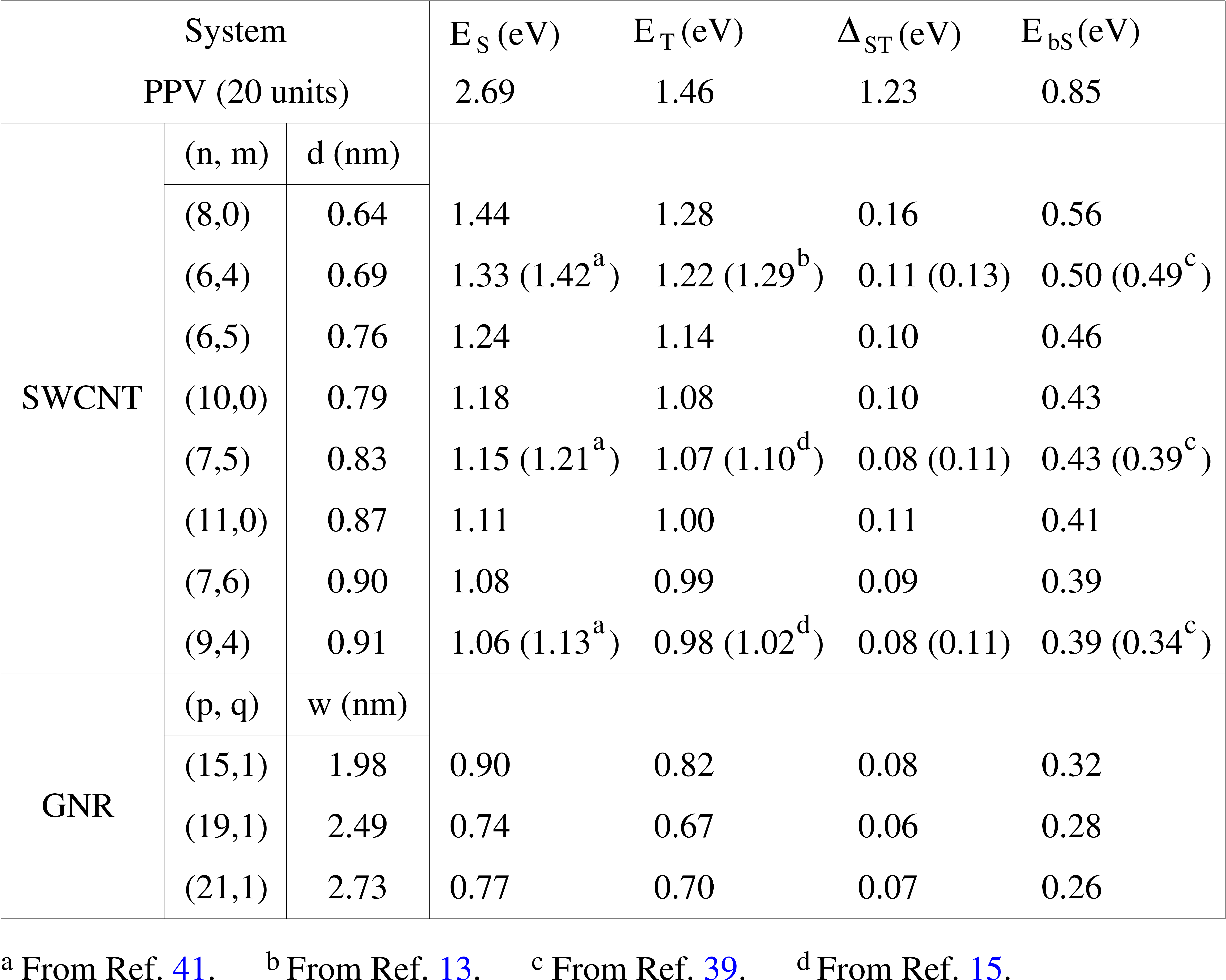}
\caption[a]{Calculated singlet and triplet exciton energies $E_S$ and $E_T$, the singlet-triplet energy difference $\Delta_{\mathrm{ST}}$, and the singlet exciton binding energy $E_{bS}$, in PPV, S-SWCNTs and AGNRs. Experimental values from Refs.~\onlinecite{Harutyunyan09a}~,\onlinecite{Matsunaga10a}~,\onlinecite{Dukovic05a} and \onlinecite{Bachilo02a} are given in parentheses.}
\label{table1} 
\end{table}
\begin{table} 
\includegraphics[width=3.2in]{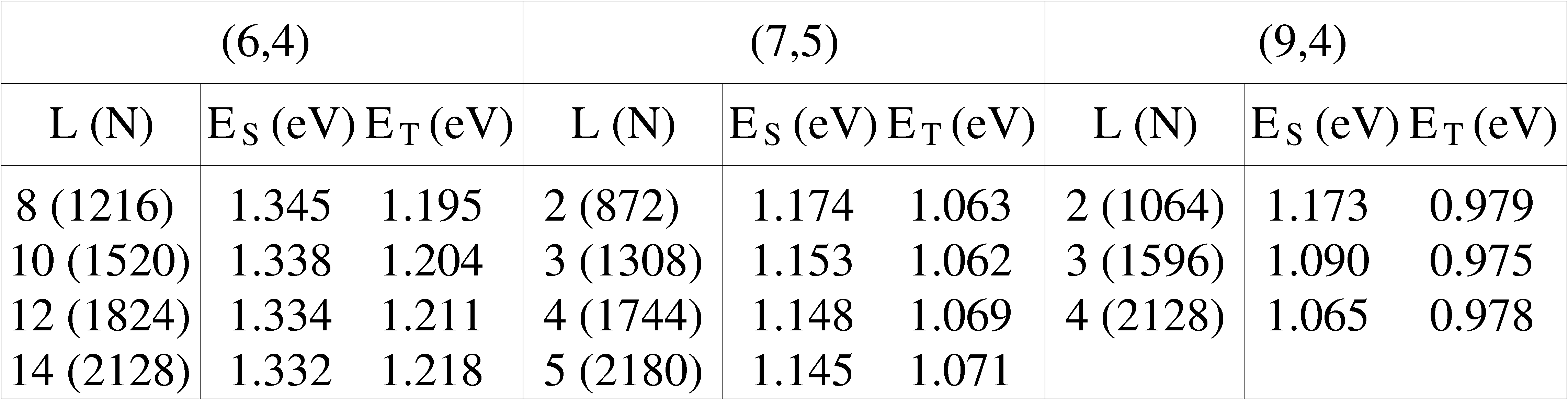}
\caption[a]{Convergence behavior of $E_S$ and $E_T$ for $(6,4)$, $(7,5)$ and $(9,4)$ S-SWCNTs with increasing length. Here $L$ ($N$) is the number of unit cells (total number of carbon atoms).}
\label{table2} 
\end{table}
\par For each of the chiral S-SWCNTs in Table~\ref{table1}, our results are for the longest nanotubes that were computationally accessible. Whether or not convergence in energies has been reached is clearly an important issue, both from the perspective of comparisons to experiments and because of the seemingly surprising difference between S-SWCNTs and PPV in the triplet sector, given that in the singlet sector $E_{bS}$ in S-SWCNTs is smaller than that of PPV by only about a factor of $2$. We have therefore done careful finite-size analysis for the three S-SWCNTs for which experimental results are available for both the singlet bright exciton and the triplet exciton. In Table~\ref{table2}, we have shown these results. Convergence to a few percent or less in both $E_S$ and $E_T$ gives us confidence about our results in Table~\ref{table1}. Comparing our results for the $(7,6)$ S-SWCNT with that in Ref.~\onlinecite{Tretiak07a}, we see that the reason for the large $\Delta_{\mathrm{ST}}$ found in that work lied not in the limitation to two unit cells, but the relatively large difference between the calculated spin-singlet exciton energy there and the experimental value.
\par
\begin{figure} 
\includegraphics[width=2.9in]{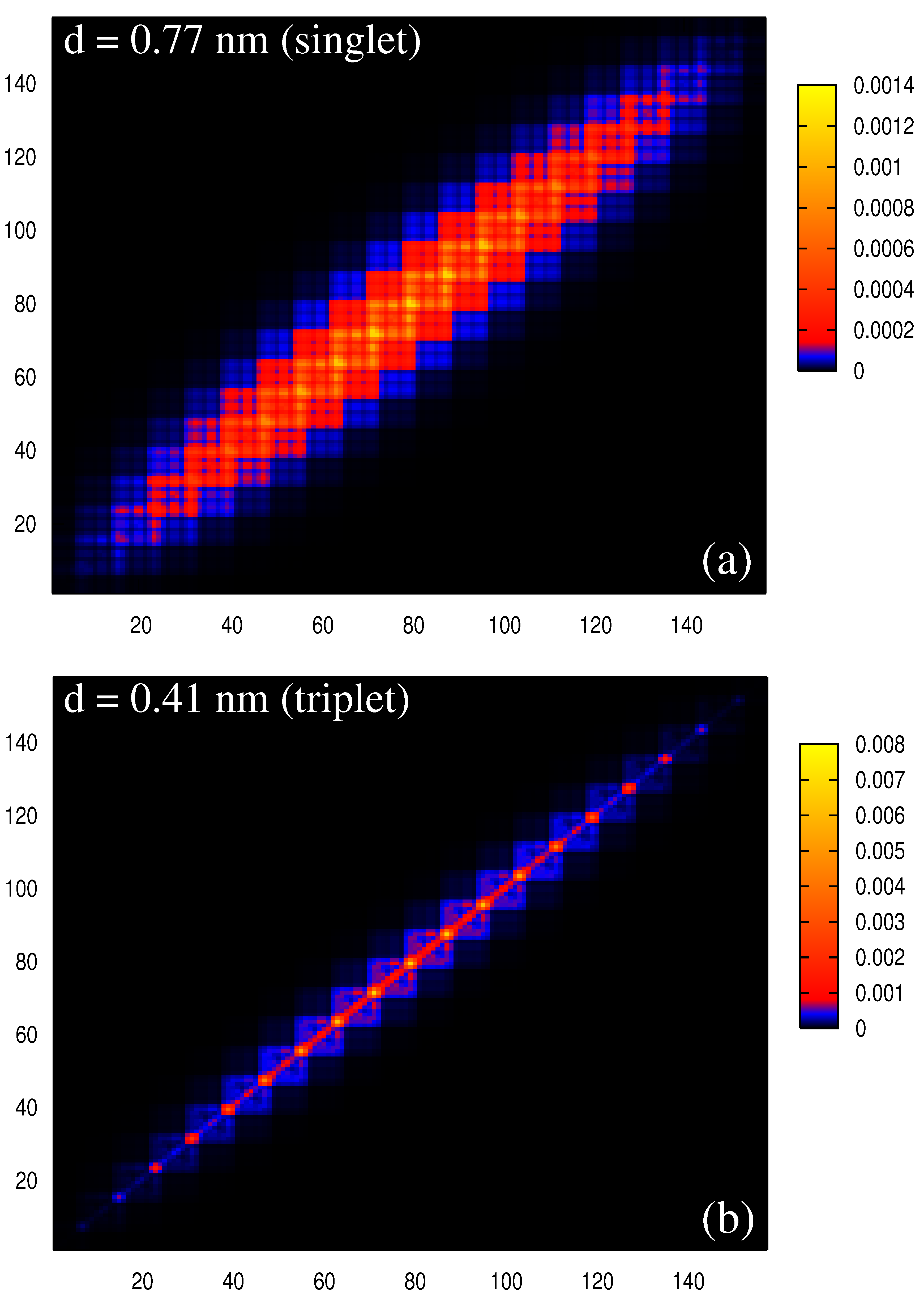}
\caption{(Color online) Probability density of the electron-hole separation (see text) in the (a) singlet and (b) triplet excitons of a 20-unit PPV chain (158 atoms) the atomic site indices. The rms electron-hole separation $d$ is included in each case. 
}
\label{f1} 
\end{figure}

 In order to understand physically the origin of the difference between PPV on the one hand and S-SWCNTS and AGNRs on the other, we have done careful wave function analyses. The amplitude for an electron located at site $i$ in real space with a hole located at site $j$ within a bound electron-hole exciton is
\begin{eqnarray}
A^{S(T)}_{ij}=\big<\Psi_{S(T)}\big|c_{i\uparrow}^\dagger c_{j\downarrow}\mp c_{i\downarrow}^\dagger c_{j\uparrow}\big|G\big>
\label{wavefunction}
\end{eqnarray}
with $|\Psi_{S(T)}\rangle$ the SCI singlet (triplet) exciton wave function and $|G\rangle$ the HF ground-state wave function \cite{Tretiak07a,Kohler04a}. In Eq.~(\ref{wavefunction}), the minus (plus) sign corresponds to the singlet (triplet) exciton. Figure~\ref{f1} presents our results for the electron-hole spatial separation probability $\rho^{S(T)}_{ij}=\big|A^{S(T)}_{ij}\big|^{2}$ plotted versus carbon atoms $i$ and $j$ for the excitons of a 20-unit PPV chain ($158$ carbon atoms). Each point $i$ on the diagonal gives the probability of finding the electron and the hole on the same location, and vertical displacements from the diagonal points $i$ gives the probability of finding a hole at some $j \neq i$, with the electron fixed at $i$. Giving multiple $i$ becomes necessary because of the open boundary condition we have chosen. Microscopic details such as electron-hole separations within a unit cell cannot be obtained from the figure, which, however, does give a coarse-grained comparison between singlet and triplet excitons.  
\par Figure~\ref{f1} indicates that the triplet exciton in PPV exhibits a significantly more localized probability density compared to the singlet exciton. This result has also been obtained previously by other investigators \cite{Beljonne96a}. For the triplet exciton in Fig.~\ref{f1}(b), the ``hot spots'' (atoms with high probability intensity colored in yellow and red according to the calibration column next to every density plot) are almost entirely along the diagonal, while for the singlet exciton in Fig.~\ref{f1}(a), hot spots are evenly distributed within the neighborhood of nearly 20 atoms about the diagonal. The root mean square (rms) electron-hole separation (``exciton size'') $d_{S(T)}=\sqrt{\sum_{ij}\rho^{S(T)}_{ij}||\vec{R_{i}}-\vec{R_{j}}||^2}$, where $\vec{R_{i}}$ ($\vec{R_{j}}$) gives the location of the electron (hole), is thus significantly different for singlet and triplet excitons ($d_{T}/d_{S}\sim 0.5$, see Fig.~\ref{f1}). This {\it qualitative} difference between the spin singlet and triplet excitons in one dimension within Eq.~(\ref{Ham}) is to be expected. We postpone further discussion until later.
\begin{figure*} 
\includegraphics[width=7.0in]{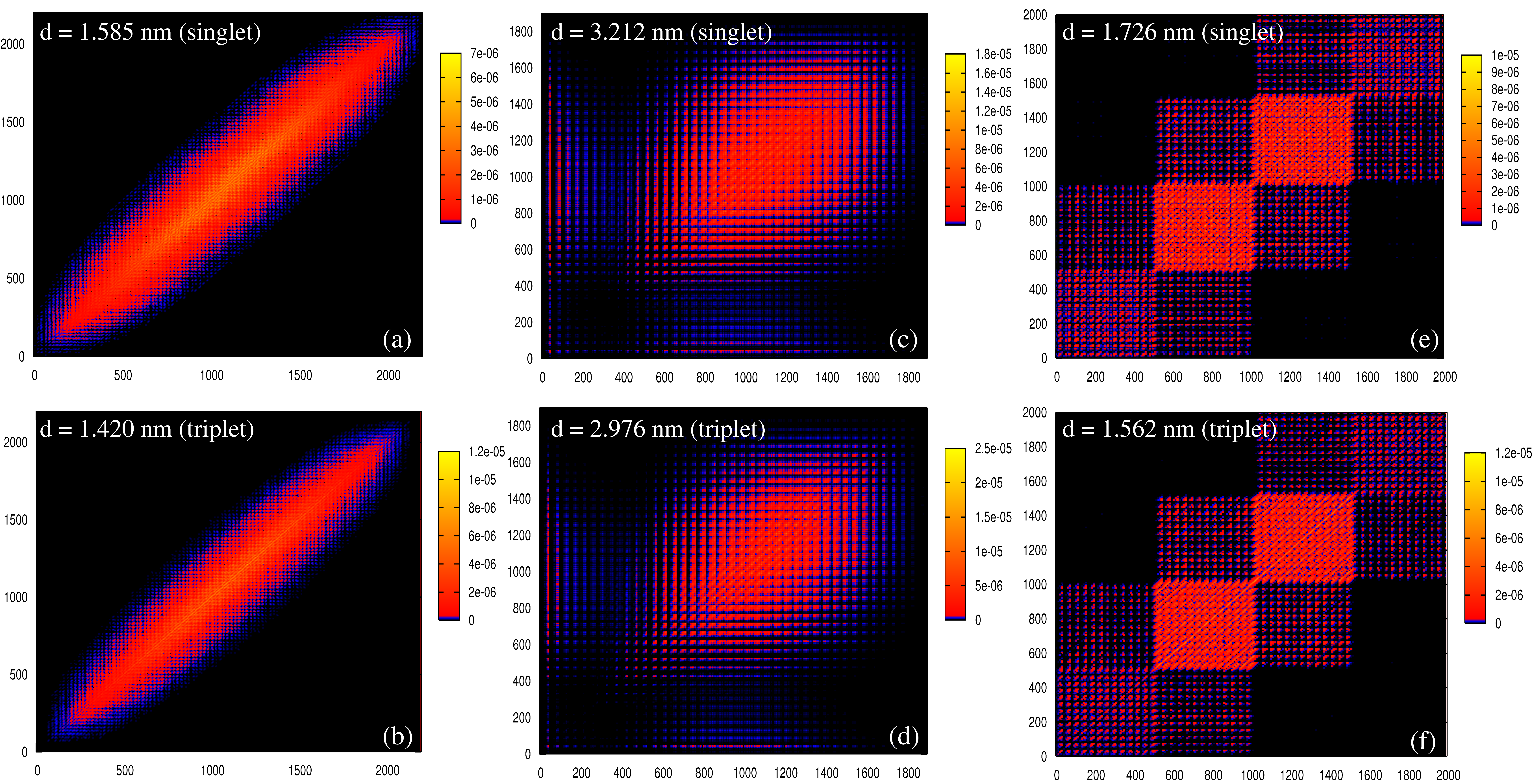}
\caption{(Color online) Probability density of the electron-hole separation in the lowest optical singlet exciton in (a) (11,0) S-SWCNT ($50$ unit cells, 2200 atoms), (c) (21,1) AGNR ($40$ unit cells, 1962 atoms) and (e) (7,6) S-SWCNT ($4$ unit cells, 2032 atoms). The probability densities of the electron-hole separation in the corresponding lowest triplet exciton are shown in (b), (d), and (f), respectively.}
\label{f2} 
\end{figure*}
\par In Fig.~\ref{f2}, we show plots of $\rho^{S(T)}_{ij}$ for the (11,0) S-SWCNT, the (21,1) AGNR, and the (7,6) S-SWCNT. The (21,1) AGNR is obtained by ``unrolling'' the (11,0) NT, while the (7,6) NT has a diameter close to that of the (11,0) NT. These $\rho^{S(T)}_{ij}$ plots are representative for all other NTs and AGNRs, respectively. The probability density patterns are wider for all three carbon nanostructures compared to PPV, and are nearly the same for spin-singlet and triplet excitons. Both of these indicate some fundamental difference between the wave functions, in particular the triplet wave functions, in $\pi$-conjugated polymers versus the carbon nanostructures. The wider probability dispersion in the AGNR indicates greater electron-hole separation in the exciton wave functions here, presumably a consequence of partial two-dimensionality, as the exciton acquires greater width within the same unit cell. The $\rho^{S(T)}_{ij}$ for the (7,6) NT, with very large unit cells is particularly interesting. Each bright square within the checkerboard pattern in Figures~\ref{f2}(e) and (f) corresponds to $i$ and $j$ within the same unit cell. The overall pattern indicates exciton wave functions delocalized over the cylinder within a unit cell, even as the unit cells themselves are coupled one dimensionally (justifying partly the characterization of NTs as ``quasi-one-dimensional''). Our results for {\it other chiral S-SWCNTs} also exhibit similar checkerboard patterns within the unit cell  for $\rho^{S(T)}_{ij}$. Similar structure should  be true also in the (11,0) zigzag NT, except that because of the much smaller unit cell size - 44 carbon atoms - this is not visible on the scale of Fig.~\ref{f2}. The wider dispersions of probability densities are consistent with the the large $d_{S(T)}$ indicated in the figures. Our calculated $d_{S}\sim 1.5$ nm for the (6,5) NT is in excellent agreement with the experimental estimate of $2.0\pm 0.7$ nm \cite{Luer08a}, giving us confidence in the accuracy of our method.
\par
\begin{figure} 
\includegraphics[width=2.9in]{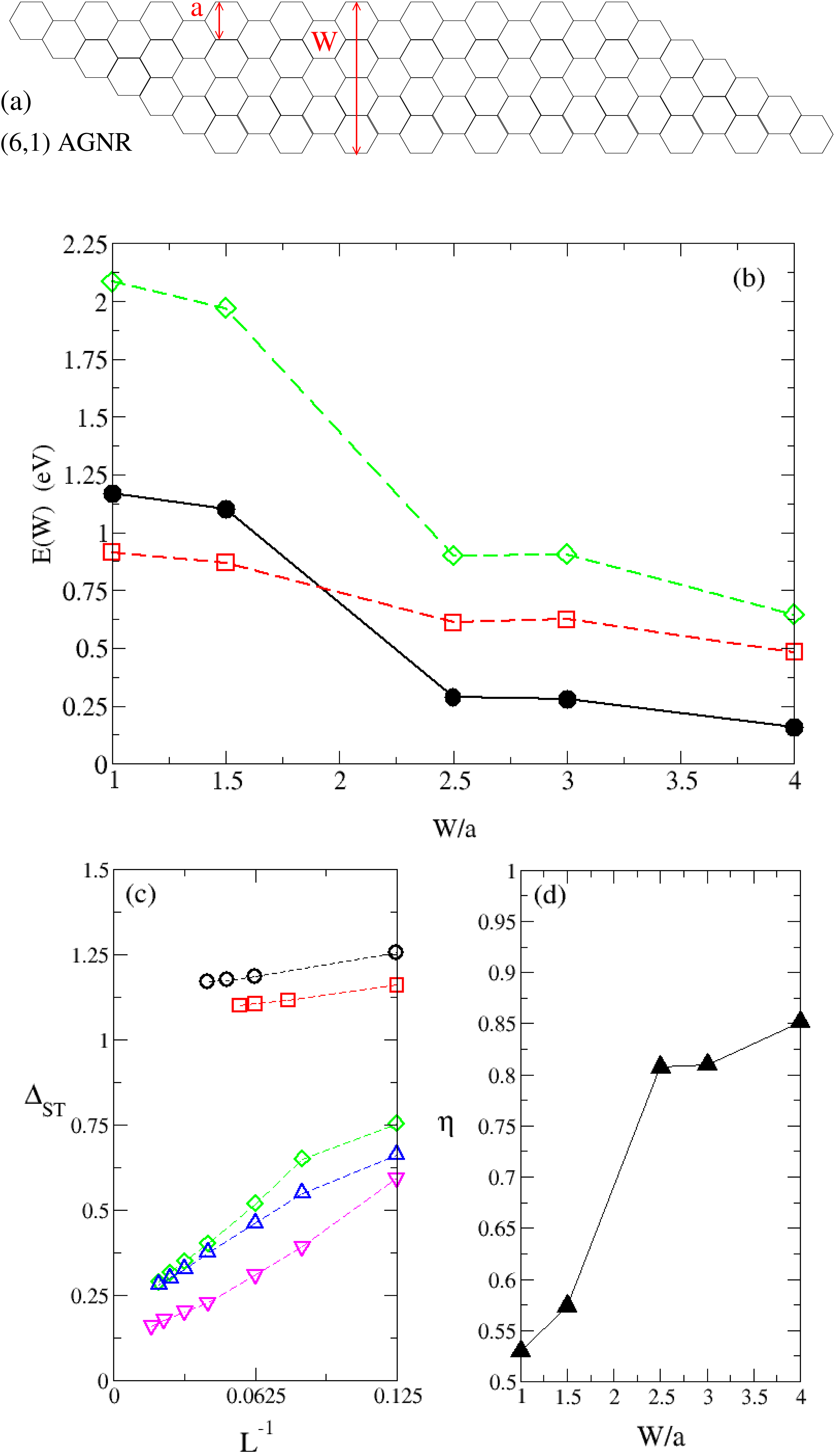}
\caption{(Color online) (a) Fragment of a (6,1) AGNR with width $W=4a$, where $a=0.249$ nm is the graphene unit lattice vector. (b) $\Delta_{\mathrm{ST}}$ (filled circle), $E_{bS}$ (open square), and $E_{bT}$ (open diamond) as functions of $W$. (c) $\Delta_{\mathrm{ST}}$ vs the inverse of the number of unit cells $L$ for PPP (circle), (1,1) (square), (3,1) (diamond), (4,1) (triangle up), and (6,1) (triangle down) AGNRs. (d) Ratio $\eta=d_{T}/d_{S}$ as a function of $W$.}
\label{f3} 
\end{figure}

 The confined versus delocalized behavior of excitons is clearly a dimensional effect. In order to understand this dimensional effect physically, we have performed systematic calculations of $E_S$ and $E_T$ for narrow AGNRs as functions of their widths (see Fig.~\ref{f3}). We start with a poly(p-phenylene) (PPP) chain as the narrowest AGNR with width $W=a=0.249$ nm, where $a$ is the graphene unit lattice vector [Fig.~\ref{f3}(a)]. By adding more PPP chains on top of one another, we generate wider (1,1), (3,1), (4,1), and (6,1) AGNRs [(2,1) and (5,1) AGNRs are ignored as they are metallic]. A large decrease in $\Delta_{\mathrm{ST}}$, by $\sim 0.9$ eV between (1,1) and (3,1) AGNRs [see Fig.~\ref{f3}(b)] indicates the emergence of quasi-two-dimensional behavior already for the (3,1) AGNR. Interestingly, $E_{bS}$ does not exhibit the sharp drop seen with $\Delta_{\mathrm{ST}}$ and $E_{bT}$. Figure~\ref{f3}(c) shows the behavior of $\Delta_{\mathrm{ST}}$ plotted against the inverse of the number of unit cells $L$ in AGNRs. As seen here, $\Delta_{\mathrm{ST}}$ has almost converged already at the smallest $L=8$ for PPP and the (1,1) AGNR. For the wider AGNRs, however, $\Delta_{\mathrm{ST}}$ continues to decrease with ribbon length even at the largest $L$ for which we have done our calculations ($L=60$). Figure~\ref{f3}(d) shows the corresponding behavior for $\eta=d_{T}/d_{S}$ as a function of $W$.
\par Let us now try to understand our main result, viz., $\Delta_{\mathrm{ST}} \sim 0.1$ eV in S-SWCNTs with $d \sim 1$ nm. $\Delta_{\mathrm{ST}}$s even smaller than ours for S-SWCNTs of similar diameters were calculated in Refs.~\onlinecite{Perebeinos05a} and \onlinecite{Capaz07a}. The smaller $\Delta_{\mathrm{ST}}$ was justified on the basis of the argument that it should be comparable to the energy splitting between optically bright and dark singlet longitudinal excitons, which is, indeed, smaller than our $\Delta_{\mathrm{ST}}$ \cite{Capaz07a,Spataru05a}. These authors claimed that the bright-dark spin singlet exciton splitting is determined by the same exchange energy that gives the singlet-triplet splitting, and hence the two quantities should be comparable. This argument is not correct however. Bright-dark exciton splitting requires {\it pairs} of occupied orbitals (labeled, say, 1 and 2) and unoccupied orbitals (1$^\prime$ and 2$^\prime$), such that within one-electron theory the optically allowed excited states $|1 \to 1^\prime \rangle$ and $|2 \to 2^\prime \rangle$ are nearly degenerate. In the presence of many-electron interactions that give nonzero matrix elements between these excited configurations new eigenstates $|1 \to 1^\prime \rangle \pm |2 \to 2^\prime \rangle$ are obtained of which one is optically bright and the other dark \cite{Zhao04a}. Thus unlike in the exchange process where only two orbitals and two electrons are involved, permutation of the coordinates of pairs of electrons are not involved in the splitting behind the bright and dark excitons. The latter is thus driven by direct Coulomb interactions, as has also been emphasized by others \cite{Chang06a}. Both the direct and the exchange Coulomb interactions vary as $1/R$, where $R$ is the exciton radius. The bright-dark splitting is small for the longitudinal excitons with large radii, but the same mechanism is behind the splitting of bright and dark {\it transverse} excitons in S-SWCNTs \cite{Wang07a} and PPV \cite{Chandross97a}, where because of the short radii of the transverse excitons, the Coulomb matrix elements are large and so are the splittings. The above argument justifies $\Delta_{\mathrm{ST}}$ larger than the singlet bright-dark separation, as the exciton radii are smaller for  the triplet than for the singlet exciton. In view of this and the results of Table~\ref{table1}, we then arrive at the following interpretation of the experiments of Ref.~\onlinecite{Kiowski07a}: the optically dark states $\sim 40$ meV below the bright exciton are spin singlet, while the ones $\sim 100$ below are spin triplets. This interpretation agrees with our results for the triplets in Tables~\ref{table1} and \ref{table2}.

\par Understanding the small $\Delta_{\mathrm{ST}}$ in spite of the relatively large $E_{bS}$, in comparison to PPV, requires going through a somewhat more complicated discussion. The PPP Hamiltonian reduces in the strong-coupling limit to a Hubbard Hamiltonian [$V_{ij}=0$ limit of Eq.~(\ref{Ham})] with effective $U$, $U_{\mathrm{eff}}\simeq U-V_{12}$, where $V_{12}$ is the nearest-neighbor repulsion. The correlated-electron wave function of $\pi$-conjugated polymers corresponds to that of a Mott-Hubbard semiconductor, since in one dimension the critical $U$ (hereafter $U_c$) for metal-insulator transition within the simple Hubbard model is arbitrarily small \cite{Lieb68a}. This implies that the ground state is dominated by covalent valence bond diagrams with only singly occupied sites \cite{Ramasesha84a}. {\it The important point now is that the nature of the ground-state wave function and of the lowest spin triplet states are related.} Thus the lowest triplet states in one dimension are also covalent, while the optical state consists of ionic valence bond diagrams \cite{Ramasesha84a}. This is why of course $\Delta_{\mathrm{ST}}$ in $\pi$-conjugated polymers is large. 
\par In contrast to one dimension, recent theoretical work has shown that in the graphene lattice $U_c$ is rather large, $U_c \geq 4t_{ij}$ \cite{Meng10a}. Graphene is a semimetal since $U_{\mathrm{eff}}$ within the PPP model is smaller than this $U_c$. It is likely that $U_c$ becomes nonzero for nanoribbons with finite width $W_c$, with $U_c$ rising gradually with width $W>W_c$. For nanoribbons that are metallic within one-electron theory and have $W>W_c$ then the ground-state wave function is no longer covalent but has strong admixing with ionic configurations with double occupancies. We {\it speculate} that a similar effect occurs in {\it semiconducting} nanoribbons with $W>W_c$, where the ground-state wave function begins to resemble that of a conventional semiconductor rather than a Mott-Hubbard semiconductor. This would also mean that the ground state is no longer covalent; furthermore, in conventional semiconductors, the singlet optical state and the triplet are both ionic and $\Delta_{\mathrm{ST}}$ is small. Our calculations in Fig.~\ref{f3}(b) then suggest that $W_c \sim 2.5a$. Although we have presented the above arguments above for the AGNRs only, it is clear that they also apply to S-SWCNTs. The remaining problem now is to understand the large $E_{bS}$ in the S-SWCNTs and AGNRs in spite of small $\Delta_{\mathrm{ST}}$.
\begin{figure} 
\includegraphics[width=3.3in]{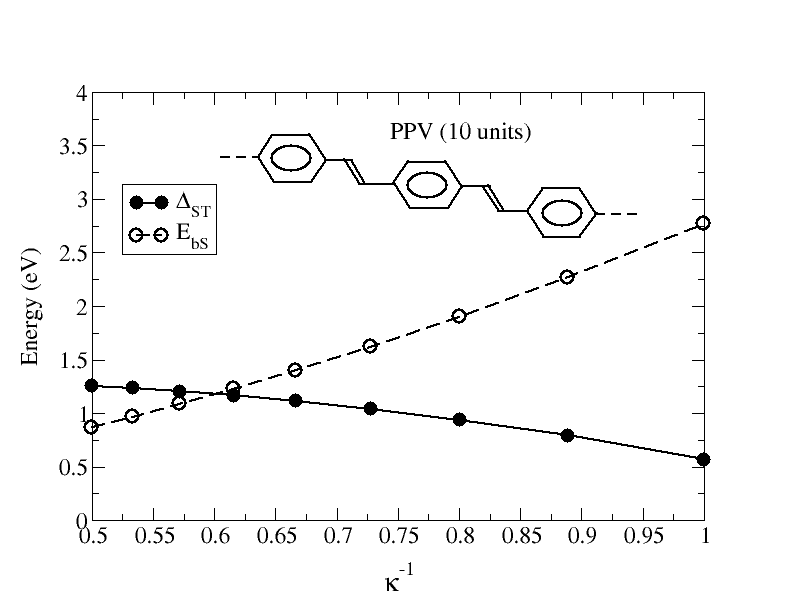}
\caption{ $\Delta_{\mathrm{ST}}$ (solid line with filled circle) vs $E_{bS}$ (dashed line with open circle) as a function of $\kappa^{-1}$ for a PPV chain with $10$ units.}
\label{f4} 
\end{figure}
\par The exciton binding energy unlike $\Delta_{\mathrm{ST}}$ is largely unrelated to the nature of the ground state, as both the exciton and the electron-hole continuum states are ionic within valence bond theory for all $U_{\mathrm{eff}}$. Hence $E_{bS}$ is given by the intersite interaction $V_{ij}$ between a particle (double occupancy) and a hole (vacancy) within the effective Hubbard Hamiltonian, and small $\Delta_{\mathrm{ST}}$ does not preclude large $E_{bS}$. A direct demonstration of all of the above (especially the existence of a $W_c$) is beyond the scope of our current work. However, large $E_{bS}$ in spite of small $\Delta_{\mathrm{ST}}$ within Eq.~(\ref{Ham}) can be demonstrated easily. In Fig.~\ref{f4}, we show our calculated $E_{bS}$ and $\Delta_{\mathrm{ST}}$ within Eq.~(\ref{Ham}) for a hypothetical PPV chain with fixed $U$ but varying $\kappa$. Smaller $\kappa$ implies larger $V_{ij}$ and smaller $U_{\mathrm{eff}}$. The behavior of $E_{bS}$ and $\Delta_{\mathrm{ST}}$ against $\kappa$ are exactly opposite, $E_{bS}$ increases while $\Delta_{\mathrm{ST}}$ decreases with increasing  $V_{ij}$.
\section{Conclusions}
\label{discussions}
\par To summarize, beyond a critical diameter in S-SWCNTs and width in AGNRs, $\Delta_{\mathrm{ST}}$ decreases rapidly with 
further increase in diameter or width. Our results for $\Delta_{\mathrm{ST}}$ are intermediate between  the very small values in Refs.~\onlinecite{Perebeinos05a} and \onlinecite{Capaz07a} and the much larger values in 
Ref.~\onlinecite{Tretiak07a}. In S-SWCNTs with $d \sim 1$ nm, and in AGNRs with comparable widths, 
$\Delta_{\mathrm{ST}}$ is nearly an order of magnitude smaller than in one-dimensional conjugated polymers. 
$\Delta_{\mathrm{ST}}$ in S-SWCNTs is smaller than the splitting between bright and dark spin-singlet excitons.
Furthermore, there is no one-to-one correspondence between the exciton binding energy and the singlet-triplet
splitting, and small $\Delta_{\mathrm{ST}}$ does not preclude moderately large exciton binding energy.
\par The small $\Delta_{\mathrm{ST}}$ in wide S-SWCNTs and AGNRs may have interesting practical applications in the area of 
organic photovoltaics. Carbon-based excitonic solar cells consist of type II heterostructures with donor and acceptor
molecular systems. Optical excitation of the donor, followed by photoinduced charge-transfer from the donor to the
acceptor, lie at the heart of the device operation here. The exciton diffusion wavelength is usually quite short,
and so is the singlet exciton lifetime. As a consequence, there is considerable loss in efficiency due to the singlet
exciton relaxing to the ground state before it reaches the donor-acceptor interface. One way to enhance the performance
is to utilize the spin-triplet exciton with a longer lifetime than the singlet exciton. Our theoretical results for
 $\Delta_{\mathrm{ST}}$ in carbon nanostructures here are interesting in this context, since
ISC in systems with small $\Delta_{\mathrm{ST}}$ is expected to be rapid and efficient. 
Enhanced photoinduced charge-transfer from the spin-triplet exciton in S-SWCNTs and AGNRs to acceptor molecules is 
therefore a distinct possibility. 
While early attempts to construct solar cells with S-SWCNTs as acceptors did not lead to encouraging 
results \cite{Kazaoui05a,Kymakis05a,Sgobba06a,Hasobe06a,Kanai08a,Landi05a}, impressive performance in blends of 
S-SWCNTs with $d \sim 1$ nm and C$_{60}$, with S-SWCNTs as the donor materials has been 
obtained \cite{Arnold09a,Bindl11a}. 
Spin-triplet participation may be behind this dramatic improvement.
\section{Acknowledgments}
\label{acknowledge}
\par We are grateful to S.~Tretiak for useful discussions. This work was partially supported by US NSF grant DMR-0705163 (K.~Aryanpour and S.~Mazumdar) and NSFC-11074077 (H.~Zhao).

\begin{thebibliography}{3}
%
\bibitem{JorioBook} A.~Jorio, G.~Dresselhaus, and M.~S.~Dresselhaus, {\it Carbon Nanotubes: Advanced Topics in the Synthesis, Structure, Properties and Applications} (Springer-Verlag, New York, LLC, 2008).
%
\bibitem{Perebeinos05a} V.~Perebeinos, J.~Tersoff, and P.~Avouris, Nano\ Lett. {\bf 5,} 2495 (2005).
%
\bibitem{Capaz07a} R.~B.~Capaz, C.~D.~Spataru, S.~Ismail-Beigi, and S.~G.~Louie, Phys.\ Stat.\ Solidi (b) {\bf 244,} 5016 (2007). 
%
\bibitem{Tretiak07a} S.~Tretiak, Nano\ Lett. {\bf 7,} 2201 (2007).
%
\bibitem{Mohite09a} A. D. Mohite, T. S. Santos, J. S. Moodera and B. W. Alphenaar, Nat. Nanotechnology {\bf 4}, 425 (2009).

\bibitem{Mortimer07a} I.~B.~Mortimer, L.~J.~Li, R.~A.~Taylor, G.~L.~J.~A.~Rikken, O.~Portugall, and R.~J.~Nicholas, \ Phys.\ Rev.\ B {\bf 76,} 085404 (2007).
%
\bibitem{Mortimer07b} I.~B.~Mortimer, and R.~J.~Nicholas, \ Phys.\ Rev.\ Lett. {\bf 98,} 027404 (2007).
%
\bibitem{Shaver07a} J.~Shaver, J.~Kono, O.~Portugall, V.~Krstic, G.~L.~J.~A.~Rikken, Y.~Miyauchi, S.~Maruyama, and V.~Perebeinos, \ Nano.\ Lett. {\bf 7,} (7) 1851 (2007).
%
\bibitem{Kiowski07a} O.~Kiowski, K.~Arnold, S.~Lebedkin, F.~Hennrich, and M.~M.~Kappes, \ Phys.\ Rev.\ Lett. {\bf 99,} (23) 237402 (2007).
%
\bibitem{Srivastava08a} A.~Srivastava, H.~Htoon, V.~I.~Klimov, and J.~Kono, \ Phys.\ Rev.\ Lett. {\bf 101,} (8) 087402 (2008).
%
\bibitem{Matsunaga08a}  R.~Matsunaga, K.~Matsuda, and Y.~Kanemitsu, \ Phys.\ Rev.\ Lett. {\bf 101,} (14) 147404 (2008).
%
\bibitem{Matsunaga09a}  R.~Matsunaga, K.~Matsuda, and Y.~Kanemitsu, \ J.\ lumin. {\bf 129,} (12) 1702 (2009).
%
\bibitem{Harutyunyan09a} H.~Harutyunyan, T.~Gokus, A.~A.~Green, M.~C.~Hersam, M. Allegrini, and A.~Hartschuh, Nano\ Lett. {\bf 9,} 2010 (2009).
%
\bibitem{Nagatsu10a} K. Nagatsu, S. Chiashi, S. Konabe and Y. Homma, Phys. Rev. Lett. {\bf 105}, 157403 (2010).

\bibitem{Matsunaga10a} R.~Matsunaga, K.~Matsuda, and Y.~Kanemitsu, \ Phys.\ Rev.\ B {\bf 81,} 033401 (2010).

\bibitem{Yang07a} L.~Yang, M.~L.~Cohen, and S.~G.~Louie, Nano\ Lett. {\bf 7,} 3112 (2007).
%
\bibitem{Yang07b} L.~Yang, C.-H.~Park, Y.-W.~Son, M.~L.~Cohen, and  S.~G.~Louie, Phys.\ Rev.\ lett. {\bf 99,} 186801 (2007).
%
\bibitem{Prezzi08a} D.~Prezzi, D.~Varsano, A.~Ruini, A.~Marini, and E.~Molinari, Phys.\ Rev.\ B. {\bf 77,} 041404(R) (2008).
%
\bibitem{Liao08a} W.~H.~Liao, G.~G.~Zhou, and F.~Xi, J.\ Appl.\ Phys. {\bf 104,} 126105 (2008).
%
\bibitem{Gundra11a} K.~Gundra, and A.~Shukla, Phys.\ Rev.\ B. {\bf 83,} 075413 (2011). 
%
\bibitem{Cao99a} Y.~Cao, I.~D.~Parker, G.~Yu, C.~Zhang, and A.~J.~Heeger, Nature (London) {\bf 397,} 414 (1999).
%
\bibitem{Kim00a} J.-S.~Kim, K.~H.~P.~Ho, N.~C.~Greenham, and R.~H.~Friend, J.\ Appl.\ Phys. {\bf 88,} 1073 (2000).
%
\bibitem{Wohlgenannt01a} M.~Wohlgenannt, K.~Tandon, S.~Mazumdar, S.~Ramasesha, and Z.~V.~Vardeny, Nature (London) {\bf 409,} 494 (2001).
%
\bibitem{Tandon03a} K.~Tandon, S.~Ramasesha and S.~Mazumdar, Phys.\ Rev.\ B {\bf 67,} 045109 (2003).
%
\bibitem{Dhoot02a} A.~S.~Dhoot, D.~S.~Ginger, D.~Beljonne, Z.~Shuai, N.~C.~Greenham, Chem.\ Phys.\ Lett. {\bf 360,} 195 (2002).
%
\bibitem{Segal03a} M.~Segal, M.~A.~Baldo, R.~J.~Holmes, S.~R.~Forrest, and Z.~G.~Soos, \ Phys.\ Rev.\ B {\bf 68,} 075211 (2003).
%
\bibitem{Carvelli11a} M.~Carvelli, R.~A.~J.~Janssen, and R.~Coehoorn, \ Phys.\ Rev.\ B {\bf 83,} 075203 (2011).
%
\bibitem{Arif09a} M.~Arif, K.~Yang, K.~Li, P.~Yu, S.~Guha, S.~Gangopadhyay, M.~F\"oster, and U.~Scherf, Appl.\ Phys.\ Lett. {\bf 94,} 063307 (2009).
%
\bibitem{Smith10a}  M.~B.~Smith, and J.~Michl, Chem. Rev. {\bf 110,} 6891 (2010).
%
\bibitem{Monkman99a} A.~P.~Monkman, H.~D.~Burrows, L.~J.~Hartwell, M.~da~G.~Miguel, I.~Hamblett, and S.~Navaratnam, Chem.\ Phys.\ Lett. {\bf 307,} 303 (1999).
%
\bibitem{Pariser53a} R.~Pariser, and R.~G.~Parr, J.\ Chem.\ Phys. {\bf 21}, 466 (1953).
%
\bibitem{Pople53a} J.~A.~Pople, Trans.\ Faraday\ Soc. {\bf 49,} 1375 (1953).
%
\bibitem{Chandross97a} M.~Chandross, and S.~Mazumdar, Phys.\ Rev.\ B. {\bf 55,} 1497 (1997). 
%
\bibitem{Wang06a} Z.~Wang, H.~Zhao, and S.~Mazumdar, Phys.\ Rev.\ B. {\bf 74,} 195406 (2006). 
%
\bibitem{Wang07a} Z.~Wang, H.~Zhao, and S.~Mazumdar, Phys.\ Rev.\ B. {\bf 76,} 115431 (2007). 
%
\bibitem{Ohno64a} K.~Ohno, Theor.\ Chim.\ Acta. {\bf 2,} 219 (1964). 
%
\bibitem{Ezawa06a} M.~Ezawa, Phys.\ Rev.\ B. {\bf 73,} 045432 (2006).
%
\bibitem{Luer08a} L.~L\"uer, S.~Hoseinkhani, D.~Polli, J.~Crochet, T.~Hertel, and G.~Lanzani, Nat. Phys. {\bf 5,} 54 (2009).
%
\bibitem{Dukovic05a} G.~Dukovic, F.~Wang, D.~Song, M.~Y.~Sfeir, T.~F.~Heinz, and L.~E.~Brus, Nano.\ Lett. {\bf 5}, 2314 (2005). 
%
\bibitem{Zhao05a} H.~Zhao, S.~Mazumdar, C.-X.~Sheng, M.~Tong, and Z.~V.~Vardeny, Phys.\ Rev.\ B. {\bf 73}, 075403 (2006). 
%
\bibitem{Bachilo02a} S.~M.~Bachilo, M.~S.~Strano, C.~Kittrell, R.~H.~Hauge, R.~E.~Smalley, and R.~B.~Weisman, science, {\bf 298}, 2361 (2002).
%
\bibitem{Kohler04a} A.~K\"ohler, and D.~Beljonne, Adv.\ Func.\ Mater. {\bf 14,} 11 (2004). 
%
\bibitem{Beljonne96a} D.~Beljonne, J.~Cornil, R.~H.~Friend, R.~A.~J.~Janssen, and J.~L.~Br\'edas, J.\ Am.\ Chem.\ Soc. {\bf 118,} 6453 (1996).
%
\bibitem{Spataru05a} C.~D.~Spataru, S.~Ismail-Beigi, R.~B.~Capaz, and S.~G.~Louie, Phys.\ Rev.\ Lett. {\bf 95,} 247402 (2005).
%
\bibitem{Zhao04a} H.~Zhao, and S.~Mazumdar, Phys.\ Rev.\ Lett. {\bf 93,} 157402 (2004).
%
\bibitem{Chang06a} E.~Chang, D.~Prezzi, A.~Ruini, and E.~ Molinari, e-print arXiv:cond-mat/0603085v1 (unpublieshed).
%
\bibitem{Lieb68a} E.~H.~Lieb, and F.~Y.~Wu, Phys.\ Rev.\ Lett. {\bf 20,} 1445 (1968).
%
\bibitem{Ramasesha84a} S.~Ramasesha, and Z.~G.~Soos, J.\ Chem.\ Phys. {\bf 80,} 3278 (1984).
%
\bibitem{Meng10a} Z.~Y.~Meng, T.~C.~Lang, S.~Wessel, F.~F.~Assaad, and A.~Muramatsu, Nature (London) {\bf 464,} 847 (2010).
%
\bibitem{Kazaoui05a} S.~Kazaoui, N.~Minami, B.~Nalini, Y.~Kim, and  K.~Hara, J.\ Appl.\ Phys. {\bf 98,} (8) 084314 (2005).
%
\bibitem{Kymakis05a} E.~Kymakis, and  G.~A.~J.~Amaratunga, Rev.\ Adv.\ Mater.\ Sci. {\bf 10,} (4) 300 (2005).
%
\bibitem{Sgobba06a} V.~Sgobba, G.~M.~A.~Rahman, D.~M.~Guldi, N.~Jux, S.~Campidelli, and  M.~Prato, Adv.\ Mater. {\bf 18,} (17) 2264 (2006).
%
\bibitem{Hasobe06a} T.~Hasobe, S.~Fukuzumi, and P.~V.~Kamat, J.\ phys.\ chem. B {\bf 110,} (50) 25477 (2006).
%
\bibitem{Kanai08a} Y.~Kanai, and J.~C.~Grossman, Nano\ Lett. {\bf 8,} (3) 908 (2008).
%
\bibitem{Landi05a} B.~J.~Landi, R.~P.~Raffaelle, S.~L.~Castro, and S.~G.~Bailey,  Prog.\ Photovoltaics {\bf 13,} (2) 165 (2005).
%
\bibitem{Arnold09a} M.~S.~Arnold, J.~D.~Zimmerman, C.~K.~Renshaw, X.~Xu, R.~R.~Lunt, C.~M.~Austin, and S.~R.~Forrest, Nano\ Lett. {\bf 9,} 3354 (2009).
%
\bibitem{Bindl11a} D.~J.~Bindl, W.~Meng-Yin, F.~C.~Prehn, M.~S.~Arnold, Nano\ Lett. {\bf 11,} 455 (2011).
%
\end{thebibliography}
\end{document}